\documentclass{article}

\usepackage{times}
\usepackage{amssymb,amsfonts}
\usepackage[fleqn]{amsmath}
\usepackage[T1]{fontenc}
\usepackage[utf8]{inputenc}

\IfFileExists{upquote.sty}{\usepackage{upquote}}{}

\IfFileExists{microtype.sty}{%
 \usepackage{microtype}
 \UseMicrotypeSet[protrusion]{basicmath} % disable protrusion for tt fonts
}{}

\usepackage[pagebackref=true]{hyperref} %sysml

\usepackage{graphicx,grffile}
\usepackage{lipsum}
\usepackage{listings}
\usepackage{afterpage}
\usepackage{endnotes,epsfig}
\usepackage{multirow}
\usepackage{xspace}
\usepackage{tabularx}
\usepackage{ragged2e}
\usepackage{booktabs}
\usepackage{paralist}
\usepackage[american]{babel}
\usepackage[shortlabels]{enumitem}
\usepackage{courier,color,wrapfig}
\usepackage{xspace}
\usepackage{balance}
\usepackage{enumitem}
\usepackage{epstopdf}
\usepackage{multirow}
\usepackage{booktabs}
\usepackage{url}
\usepackage{amssymb}
\usepackage{pifont}
\usepackage{subfigure}
\usepackage{tikz}
\usepackage{comment}
\usepackage{mathtools}
\usepackage[normalem]{ulem}
\usepackage{xkeyval}
\usepackage[algo2e]{algorithm2e}
\usepackage{ifthen}
\usepackage{etoolbox}
\usepackage[margin=5pt,font=small,labelfont={bf,rm}]{caption}
\usepackage{bbding}
\usepackage{enumitem}

\usepackage{array}
\newcolumntype{L}[1]{>{\raggedright\let\newline\\\arraybackslash\hspace{0pt}}m{#1}}
\newcolumntype{C}[1]{>{\centering\let\newline\\\arraybackslash\hspace{0pt}}m{#1}}
\newcolumntype{R}[1]{>{\raggedleft\let\newline\\\arraybackslash\hspace{0pt}}m{#1}}
\usepackage{amsmath}
\usepackage[textsize=tiny]{todonotes}

\PassOptionsToPackage{usenames,dvipsnames}{color} % color is loaded by hyperref

\makeatletter
\def\maxwidth{\ifdim\Gin@nat@width>\linewidth\linewidth\else\Gin@nat@width\fi}
\def\maxheight{\ifdim\Gin@nat@height>\textheight\textheight\else\Gin@nat@height\fi}
\makeatother
\setkeys{Gin}{width=\maxwidth,height=\maxheight,keepaspectratio}


\providecommand\parab[1]{\noindent\textbf{#1}}
\providecommand\parae[1]{{\vspace{-1mm}\textit{#1}}}

\apptocmd\normalsize{%
\abovedisplayskip=5pt
\abovedisplayshortskip=5pt
\belowdisplayskip=5pt
\belowdisplayshortskip=5pt
}{}{}

\hypersetup{final}
\newcommand{\sysname}{\textsc{ML-EXray}\xspace}

\newcommand{\etc}{\textit{etc.}\xspace}
\newcommand{\ie}{\textit{i.e.,}\xspace}
\newcommand{\eg}{\textit{e.g.,}\xspace}

\newcommand{\secref}[1]{\S\ref{#1}}
\newcommand{\figref}[1]{Figure~\ref{#1}}
\newcommand{\tabref}[1]{Table~\ref{#1}}

\newcommand{\eqnref}[1]{Eqn.~\ref{#1}}

\newcommand{\squishlist}{
 \begin{list}{$\bullet$}
 		{ \setlength{\itemsep}{0pt}
 			\setlength{\parsep}{3pt}
 			\setlength{\topsep}{3pt}
 			\setlength{\partopsep}{0pt}
 			\setlength{\leftmargin}{1.5em}
 			\setlength{\labelwidth}{1em}
 			\setlength{\labelsep}{0.5em} } }

\newcommand{\squishend}{
  \end{list}  }

\usepackage[accepted]{mlsys2021}

\mlsystitlerunning{\sysname}

\begin{document}

\twocolumn[
\mlsystitle{\sysname: Visibility into ML Deployment on the Edge}

\begin{mlsysauthorlist}
\mlsysauthor{Hang Qiu}{to}
\mlsysauthor{Ioanna Vavelidou}{to}
\mlsysauthor{Jian Li}{goo}
\mlsysauthor{Evgenya Pergament}{to}
\mlsysauthor{Pete Warden}{goo}
\mlsysauthor{Sandeep Chinchali}{ed}
\mlsysauthor{Zain Asgar}{to}
\mlsysauthor{Sachin Katti}{to}
\end{mlsysauthorlist}

\mlsysaffiliation{to}{Department of Electrical Engineering, Stanford University, Stanford, USA}
\mlsysaffiliation{goo}{Google, Mountain View, USA}
\mlsysaffiliation{ed}{Department of Electrical and Computer Engineering, University of Texas, Austin, USA}

\mlsyscorrespondingauthor{Hang Qiu}{hangqiu@stanford.edu}

\mlsyskeywords{Edge ML, ML Deployment Validation, ML Debugging}

\vskip 0.1in

\begin{abstract}
Benefiting from expanding cloud infrastructure, today's deep neural networks (DNNs) have increasingly high performance when trained in the cloud. Researchers spend months of effort competing for an extra few percentage points of model accuracy. However, when these models are actually deployed on edge devices in practice, very often, the performance can abruptly drop over 10\% without obvious reasons. The key challenge is that there is not much visibility into ML inference execution on edge devices, and very little awareness of potential issues during the edge deployment process. We present \sysname, an end-to-end framework which provides visibility into layer-level details of the ML execution, and helps developers analyze and debug cloud-to-edge deployment issues. More often than not, the reason for sub-optimal edge performance does not only lie in the model itself, but every operation throughout the data flow and the deployment process. Evaluations show that \sysname can effectively catch deployment issues, such as pre-processing bugs, quantization issues, suboptimal kernels, \etc Using \sysname, users need to write less than 15 lines of code to fully examine the edge deployment pipeline. Eradicating these issues, \sysname can correct model performance by up to 30\%, pinpoint error-prone layers, and guide users to optimize kernel execution latency by two orders of magnitude. Code and APIs will be released as an open-source multi-lingual instrumentation library and a Python deployment validation library.
\end{abstract}

]

% \printAffiliationsAndNotice{}  % leave blank if no need to mention equal contribution
% \printAffiliationsAndNotice{\mlsysEqualContribution} % otherwise use the standard text.

\section{Introduction}
\label{sec:Intro}

Driven by continuous advances in machine learning (ML) techniques, ML-based systems have been increasingly deployed to accomplish a variety of tasks. With the rise of the Internet of Things (IoT), the endless sensory data stream from edge devices to the cloud helps train ML models to achieve higher and higher performance. In addition to sitting on the cloud serving inference requests, recent years have seen these high-performance models widely deployed back on actual devices at the edge, to enable low-latency, low-power, privacy-sensitive applications (\eg autonomous vehicles, personal assistants, video analytics \etc).

However, deploying these ML applications  on the edge poses new challenges, specifically due to compute hardware heterogeneity, environmental variations, and sensor variations~\cite{MLSYS2021_b53b3a3d}. These issues are not well addressed, partly because there is a disconnect between the people who develop the models in the cloud and those who deploy the model on edge devices. This disconnect is deeply rooted because the design goals are not exactly the same: the objective of training is heavily focused on accuracy whereas the deployment prioritizes efficiency and end-to-end system performance. There are a lot of design choices being made when people are training and improving the model in the cloud that are not well documented and are lost in the handoff to the app development and deployment team. Even if there was a good handoff, the heterogeneity at the edge can be in conflict with those design choices.

This disconnect can be critical to efficient and successful deployment. 
% to ML researchers as well as application developers. On the cloud side, researchers spend months of effort to develop lightweight models~\cite{rastegari2016xnor}, while trying to maintain, if not sacrifice a few percent of, the performance metrics. On the edge side, these models, when deployed, can have significantly degraded performance, compared to their cloud counterparts, without any obvious reasons. Very often, 
For example,  a popular internet search company has been experiencing significantly degraded performance (see \secref{sec:preprocess_bugs}) when they deploy their vision-based ML applications using popular models (\eg MobileNet~\cite{howard2017mobilenets}). A large social media company finds that some well-trained neural networks, when quantized and deployed on their app, are not working for particular chips (see \secref{sec:quant_impact}). It is very hard to pinpoint where the problem lies. More often than not, not only the operations (Ops) in the model could be error-prone when executed on heterogeneous devices, but also bugs throughout the whole pipeline (\eg preprocessing, postprocessing, model optimization, quantization).

What exacerbates the situation is that it is very challenging to debug the problems caused by this disconnect.
There is neither any visibility into, nor any reference of, what is happening when models are executed at the edge. 
Without proper deployment benchmarks, many of these issues will just slip by silently. 
Take image classification models (\eg Keras classification models~\cite{keras_app}) as an example. A MobileNet model takes an \textit{RGB} image of $[-1.0,1.0]$ as input, whereas a VGG~\cite{simonyan2014very_vgg} model takes a \text{BGR} image, 
% using OpenCV~\cite{opencv_imread}, 
and a DenseNet~\cite{Gao2017densenet} model takes $[0.0,1.0]$ inputs. These input mismatches will not trigger any run-time errors, but can affect model performance silently(~\secref{sec:preprocess_bugs}).
Even for those issues caught belatedly, due to the \textit{lack of visibility}, and the \textit{asymmetric information/assumptions} from the disconnect, it takes weeks and even months to debug.
The engineers have to manually log the output from any ops they suspect from the complicated neural network graph. Then they verify these logs against a correct pipeline, often  developed by themselves as well. The process is extremely laborious and cumbersome.

% - Today's neural nets are well trained on the cloud, with months of sweat improving model's accuracy by a few percent

% - However, when these models are actually deployed on edge devices in practice, very often, the performance is dropping over 10\% all of a sudden without obvious reasons

To provide visibility and help facilitate debugging edge ML systems, we present \sysname, a cloud-to-edge deployment validation framework. \sysname enables app developers to discover and locate possible issues faster and provide root-cause analysis when possible. Specifically, \sysname 1) scans model execution in edge ML apps by logging intermediate outputs, 2) provides a faithful replay of the same data using a reference pipeline, 3) compares performance differences and per-layer output discrepancies, and 4) enables users to define logging around custom functions, write custom reference pipelines, and write custom assertions to verify suspicious model behaviors.

% - challenges: low visibility to ml execution on the edge

% - This tool (to be crisply named) will allow you to discover why, and more often than not, the reason does not only lie in the model itself, but every operation around it, from pre-processing to post-processing.
    
% - summary of case studies and findings

% We explore several ways of using \sysname

We implement \sysname as a suite of instrumentation APIs and a Python library. 
% We develop and evaluate \sysname 
Evaluations are conducted across image, audio, and text-based ML applications using widely deployed models and public benchmark datasets. Instrumenting and deploying these applications on mobile devices, we found that \sysname can effectively catch a variety of deployment issues that the industry has been suffering, including preprocessing bugs, model optimization and quantization issues, and abnormal execution latency. Some of these issues, such as malfunctioning quantization ops, were not previously discovered, and have been reported and are now on the developer teams' road map to be fixed. To catch these issues using \sysname, users would only need to write less than 5 lines of code (LoC) of app instrumentation, and less than 10 LoC of assertion functions. The overhead of \sysname during runtime is only up to 3ms per frame (2.3\% if running the model on CPU, 15\% if on GPU), consuming up to 4 MB memory (compared to a typical ML model footprint of 15 to 100 MB). 

A major usage of \sysname is to trigger per-layer examination and check user-defined assertions to quickly and effectively catch the issues mentioned above. For example, comparing against the reference pipeline step by step, a difference  after normalization op can verify if the scaling is correct,  a difference after a quantized layer can alert a quantization issue, a comparison after permuting the channels of the preprocessing output can verify if RGB channel was arranged as BGR. 
All of the per-layer logging and assertions can be run offline efficiently.
Resolving these issues, our evaluations show that  \sysname can help users 1) significantly correct app performance (\eg improving model accuracy by up to 30\%), 2) pinpoint buggy ML operations (\eg quantized depthwise convolution layer, quantized average pooling layer), and 3) shorten execution latency.
% by using optimized kernels.

In summary,  we make the following contributions:

\vspace{-4mm}
\begin{itemize}[leftmargin=4mm]
    \item We introduce edge ML instrumentation APIs, providing visibility into layer-level details on edge devices.
    \vspace{-2mm}
    \item We provide an end-to-end edge deployment validation as an abstraction, giving users an interface to design custom assertions for deployment verification.
    \vspace{-2mm}
    \item We show that \sysname can catch various deployment issues in large-scale industrial production pipelines across tasks of different data modalities.
    \vspace{-2mm}
    \item We demonstrate the impact of these previously invisible issues, and how \sysname makes it easy to catch these issues,  raising the awareness of deployment debugging.
    
\end{itemize}
\vspace{-3.5mm}
Code and APIs will be  open-sourced to the community.
% as a multi-lingual instrumentation library and a Python deployment validation library.

% \hang{file the issues if not found on github, add links to those as evidence of impact, probably not doable as anonymous double blind policy}

% \parab{Edge ML instrumentation APIs.}

% \parab{Cloud-to-Edge deployment validation framework. }

% \parab{User Interface for custom logging and domain knowledge assertions.}
\section{ML Deployment on the Edge}
\label{sec:background}
In this section, 
we describe the Edge ML deployment process and challenges experienced by our industrial partners during their large-scale deployment. We use these critical deployment issues as case studies throughout the paper.
Deploying pre-trained ML models usually involves three major steps: 1) develop an ML inference pipeline on the target edge platform, 2) convert ML checkpoints to executable versions, and, when necessary, quantize models to fit hardware specifications, and 3) validate model performance and debug potential deployment issues.

% ML-Exray can efficiently find bugs that have actually occurred during a large company's edge deployment and testing phases. This industrial partner provided use cases and examples of critical deployment bugs that we use as case studies throughout this paper, such as XYZ
% %
% Without loss of generality, we use Tensorflow models as an example to describe these in detail as follows.

\parab{ML Inference Pipeline on the Edge.} At the core of an ML application is an inference pipeline, which applies edge device sensory data to a pre-trained ML model to perform an automated task. A typical pipeline includes sensor capture (\eg image, audio, text, point cloud, \etc), data pre-processing (\eg resizing, normalization, voxelization), invoking the ML model, and finally post-processing results. 

Take an android image classification app as an example. The app captures images from the camera, extracts RGB channels from the image byte array, resizes it to the model's input size, converts the input to the numerical format expected by the model, forwards the data to the input tensor, invokes the model, gets the output class of the highest score, and finally reports the corresponding image category. 

It is extremely challenging for app developers to write such a pipeline without any bugs, especially for the data preprocessing stage. Essentially, they have to create new code that emulates what happens in the training pipeline exactly, with no easy way to check that they're doing it correctly.

From conversations with engineers in industry responsible for shipping on-device ML products, we've identified several categories of errors that are common in real products and are currently hard to detect. These can most easily be illustrated using examples from the Android image classification example described above.

% \parae{a) Image capture.} As typical DNNs require square input images, does the app crop out a square from the center of the image or resize the entire image? 

\parae{Channel extraction.} Images are natively stored according to a channel arrangement (\eg YUV, BGR, \etc) that differs from that expected by the DNN. Bugs in the conversion process can affect the results. Even if the channel is arranged correctly, the library being used to extract the RGB values can be important, since there can be differences in color space and gamma conversions~\cite{wallace1992jpeg}.

% \hang{
% Pete, do you have citation/reference for this?
% I don't have a good academic citation, but the JPEG standard means that there can be small bit differences in the decoded image: https://stackoverflow.com/questions/23565889/jpeg-images-have-different-pixel-values-across-multiple-devices
% }

\parae{Resizing.} Camera captured images are almost certainly not the right size for the network's input, and will probably need to be scaled down.
% , which requires a choice of downsampling algorithms
In the early days, instead of an area-averaging downsampler, a lot of researchers just used a default of bilinear resampling, which can introduce a lot of aliasing~\cite{resizing_func}. 
% \hang{
% Pete, do you have citation/reference for this?
% I don't, unfortunately, but this blog post is related: https://medium.com/hackernoon/how-tensorflows-tf-image-resize-stole-60-days-of-my-life-aba5eb093f35
% }
This mismatch proved to lose a decent amount of top-1 accuracy, as we show in \secref{sec:preprocess_bugs}.

\parae{Numerical conversion.} Color values are usually expressed as unsigned integers, but DNNs usually want floating point values in the training environment. 
% That leaves a lot of choices available for converting from integers to floats - for example I've seen 0.0 to 255.0, -1.0 to +1.0, or 0.0 to 1.0. 
This conversion usually happens deep in the internals of the training framework, of which the model author is unlikely to be aware, and so isn't able to pass along the information to the mobile app developer even if they wanted to. 
% That leaves the deployment engineer to figure it out for themselves, and if they get it wrong it may not be obvious from using the app. 
It is tricky to detect such a mismatch. For example, if the network expects $[-1.0,1.0]$ and the conversion produces $[0.0, 1.0]$, it will just appear as a washed-out image and so recognition will somewhat appear to work, just with a big loss of accuracy (see \secref{sec:preprocess_bugs}).

\parae{Orientation.} The input image, either landscape or portrait, always has the correct orientation during training. However, when edge devices capture image data, the orientation can change. Some networks may be trained with data augmentation, such as random rotations or horizontal/vertical flips. Our evaluation (\secref{sec:preprocess_bugs}) shows that many popular classification models suffer from a 90 degree rotation of the input.

The same challenges also exist in apps with other input types such as audio, accelerometers, point clouds, RF signals. These sensor modalities typically require a lot more explicit feature generation in the preprocessing stage, such as FFTs~\cite{yamnet}, voxelization~\cite{zhou2018voxelnet}, \etc Usually, this feature generation work, implemented outside of the ML model graph, is not available to app developers.

% \hang{Better give example raw data and preprocessed input, maybe earlier, no space}

\parab{Model Optimization and Quantization.}
The most important module
% , yet with least ``visibility'', 
in the edge ML inference pipeline is the neural network itself. When deploying these models on edge devices, it is crucial to optimize~\cite{han2016deep, tfmodelquant} these models for inference purposes as they come typically from a training pipeline. In contrast to a training environment, edge devices often have limited compute resources and memory. Fortunately, app developers have a plethora of choices to optimize the model. For example, to invoke these pre-trained DNNs efficiently on edge devices, training-related features and operations are removed and the network is optimized with techniques such as constant folding (including batch normalization folding) and fusion (including fusion of activation function, such as ReLU). Additionally, edge ML systems usually use light-weight interpreters to replace the full TensorFlow run-time~\cite{MLSYS2021_d2ddea18}.
% \hang{add citation}
% \hang{Pete, Jian, please feel free to chime in}
% You could reference https://arxiv.org/abs/2010.08678 for the Micro interpreter discussion.

One of the most common optimizations is model \textit{quantization}~\cite{jacob2017quantization, krishnamoorthi2018quantizing}: converting model weights and biases from 32-bit floats to lower precision values (\eg 16-bit floats for GPUs, 8-bit integers for Edge TPUs). Quantization offers faster arithmetic, reduces memory, disk, network, and battery consumption, and enables running models on hardware accelerators, some of which do not support floating-point operations. Quantization has been successfully applied to many model operations~\cite{Shangguan2019Speech} and architectures~\cite{Li2021RNN}. In the most simple format (asymmetric, per-tensor quantization), the equation to quantize 32-bit floats to unsigned integers and reconstruct back is:
% \vspace{-2mm}
{\small
\begin{equation}
quantized = uint8 \left(\frac{original - min}{max - min} \times 255\right)
\label{eqn:quantize}
\end{equation}
\vspace{-4mm}
\begin{equation}
reconstructed = \frac{quantized}{255.0} \times (max - min) + min   
\label{eqn:unquantize}
\end{equation}
}
% \vspace{-1mm}
% \begin{lstlisting}[language=Java]
% quantized = uint8(((original - min) / (max - min)) * 255)
% reconstructed = ((quantized / 255.0) * (max - min)) + min
% \end{lstlisting}
where $max$ and $min$ are the profiled range of the tensor value. In reality, the details are much more complicated than the equations and we select a few outstanding reasons:

\begin{figure*}
 \centering
%  \begin{minipage}{0.9\textwidth}
\vspace{-2mm}
  \includegraphics[width=0.8\textwidth]{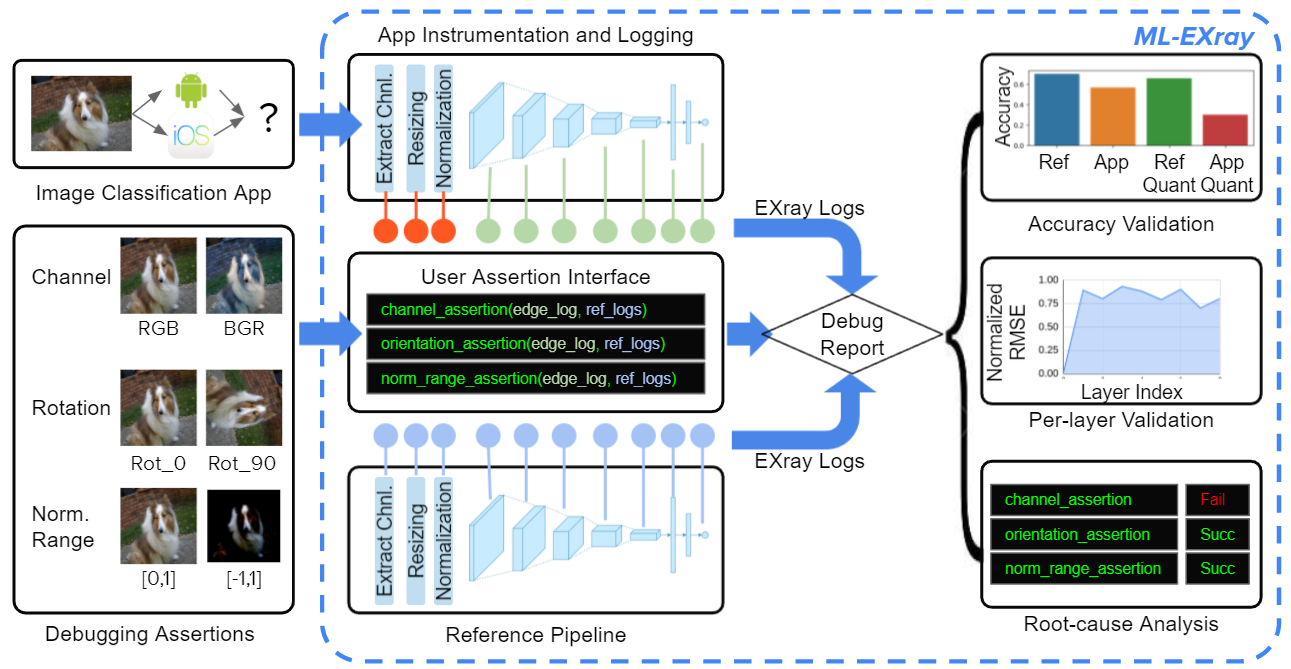}
  \vspace{-4mm}
  \caption{An Overview of the \sysname System. (1) App developers use \textit{\sysname APIs} (\secref{sec:api}) to instrument mobile apps (\eg an image classification app) to produce \textit{EXray logs} (default logs (green) and custom logs (red)). (2) \sysname plays back the same data using a reference pipeline (\secref{sec:ref_pipeline}) and produces the same user-defined logs. (3) Users define debugging assertions (\secref{sec:validation}) like input channel arrangement, orientation, and normalization range. (4) \sysname compares edge logs with reference logs (\secref{sec:validation}), and reports potential issues like accuracy degradation, per-layer output drift, as well as  root-cause assertion results.}
  \label{fig:sysarch_example}
  \vspace{-4mm}
%   \end{minipage}
\end{figure*}

\parae{Scale calibration.} The $min$ and $max$ from \eqnref{eqn:quantize}, \ref{eqn:unquantize} for each tensor is unknown to app developers. Quantization tools typically require example input data for calibration. In this case, an outlier in the representative dataset could inflate the scale such that quantized integers do not have enough resolution to tell the difference generated by normal input data. In contrast, a small example dataset could render the scale too small such that values generated by normal input data are clipped by integers 0 and 255.

\parae{Symmetric quantization.} For a given tensor, asymmetric quantization utilizes the full int8 range while symmetric quantization  uses a fraction 
% of the full int8 range 
when data is skewed. However, in practice, symmetric quantization is widely used, allows better 
optimization on ARM CPUs, and the absence of \textit{zero point} translates to better performance~\cite{nagel2021white}.
% and a tensor of range [-127, 127] allows better 
% optimization 
% for single instruction multiple data (SIMD) 
% on an ARM CPU.

\parae{Per-tensor vs per-channel quantization.} The weight distribution in a tensor plays a key role in the accuracy of the model. After batch normalization weight folding, the weight in a convolution or a fully-connected (FC) layer can sometimes be very different from channel to channel. In this case, \textit{per-tensor quantization} can squash the entire channel to 0 due to the scale difference,
% It's possible a whole channel is quantized to zero if the values are generally smaller compared with the maximum value in the weight tensor, which is equivalent of dropping one channel in the weight. 
whereas \textit{per-channel quantization} allows each channel to have its own 
% \textit{min}, \textit{max},
% hence 
\textit{scale} 
and \textit{zero point}.
% , so scale and zero points become vectors, instead of scalars.

% \hang{add quotes to github issues here}

% \hang{Pete, Jian, feel free to list other issues related to quantization here}

% \hang{cite binary net paper}

\parab{Debugging ML Deployment on the Edge.}
Depending on the severity of these issues described above, model performance can vary from somewhat working (with much lower precision) to not working at all, \ie outputting constant values (see \secref{sec:eval}).
At the same time, multiple issues can exist together, leaving app developers clueless to debug.
Moreover, even though there are run-time ML benchmarking~\cite{reddi2020mlperf_inf, reddi2020mlperf_mobile_inf} and debugging~\cite{kang2018model_assertion} tools assuming a correct deployment, there are few tools to help figure out if there is any problem lying in the deployment process. App developers would have to manually log and inspect specific tensors, \eg function outputs, quantized model weights and operation outputs, and check them with the original model given the same input. This process is extremely tedious and laborious, and demands significant deployment  experience and domain knowledge.

% \parab{Deployment Validation.}

% tensorflow lite

% java

% cpp

% edgetpu

% \input{fig_latex/sys_arch_0}

\section{\sysname}
\label{sec:design}

To provide visibility into and solve edge ML deployment issues, we propose \sysname, an edge ML deployment validation and debugging framework.

\vspace{-2mm}
\subsection{System Overview}
\vspace{-1mm}
\sysname consists of three components: 1) a cross-platform API for instrumentation and logging (\secref{sec:api}) for  ML running on the edge and cloud, 2) a reference pipeline (\secref{sec:ref_pipeline}) for data playback and establishing baselines, and 3) a deployment validation framework (\secref{sec:validation}) to identify issues and analyze root-causes. In addition, 
% to built-in functionalities, 
\sysname is fully customizable to user-defined verification.

\figref{fig:sysarch_example} shows a system overview.
We continue to use the image classification app as an example to describe the \sysname system. 
For example, to verify whether the app has any preprocessing issues, an app developer (user) can use \sysname APIs to instrument around preprocessing functions (red dots) in the ML inference pipeline. While running the app, \sysname will collect both default inference logs (green dots) as well as custom logs (red dots). In order to validate if there are any preprocessing issues, app developers can insert specific assertions that they suspect. For example, users can write simple debugging assertion functions, such as checking whether the channel arrangement is correct, whether the input orientation is correct, or whether the normalization range is correct. Taking these assertion functions, \sysname runs an instrumented reference pipeline and compares the logs. In the example shown in \figref{fig:sysarch_example}, \sysname first checks the accuracy (\textit{accuracy validation}). If there is an accuracy drop, \sysname then performs per-layer error analysis to locate the discrepancy (\textit{per-layer validation}). 
% at channel extraction (first preprocessing function). 
Finally, it runs all built-in and user-defined assertions (\textit{root-cause analysis}).

% \hang{rewrite the following para and figure 2}

% \figref{fig:sysarch} shows a system diagram. 
% A general workflow is as follows. Both the instrumented mobile app (\secref{sec:api}) and the reference pipelines (\secref{sec:ref_pipeline}) will instantiate the EdgeML Monitor (\secref{sec:data_model}) to collect telemetry data of ML inferences. \sysname uses these data to compare and validate the mobile ML pipeline against the reference pipeline (\secref{sec:validation}). For well-defined tasks, such as image classification, object detection, audio recognition, \sysname provides built-in assertions on critical error-prone points (\eg preprocessing) where bugs are common. 

\begin{figure}
 \centering
  \includegraphics[width=0.8\columnwidth]{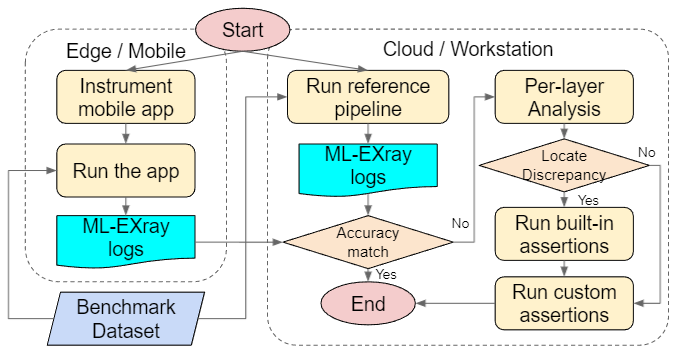}
  \vspace{-3.5mm}
  \caption{\sysname generic deployment validation flowchart. \sysname collects logs from instrumented app and reference pipelines and compares them for deployment validation: 1) match the accuracy between two pipelines, 2) If accuracy drops, scrutinize layer-level details and locate the discrepancy. 3) run assertion functions for root-cause analysis.}
  \label{fig:debuggingflowchart}
  \vspace{-4.5mm}
\end{figure}

\figref{fig:debuggingflowchart} shows a generic debugging flowchart. \sysname takes as input the model, the benchmark dataset, and produces the edge logs via instrumentation (\secref{sec:api}). It runs the reference pipeline (\secref{sec:ref_pipeline}) using the same model on the same dataset in a non-edge environment (\eg cloud). A degraded accuracy from the edge logs indicates potential deployment issues.  \sysname then compares the logs from the mobile and cloud to locate the error-prone ops (\secref{sec:validation}). 
% \sysname consists of built-in assertions corresponding to discrepancies at different stages of the pipeline to analyze the potential reason. 

For novel tasks or domain-specific verification, \sysname provides a user interface to insert domain knowledge via a) adding custom logs, b) writing custom assertion functions, c) providing user-defined reference pipelines as baselines. Taking lane detection as an example, users can add lane location to logs, define lane distance as assertions, and provide special post-processing functions to the inference and reference pipelines. Finally, \sysname runs user-defined assertions to incorporate expert domain knowledge for root cause analysis.
Next, we detail the three core components.
% of \sysname in this section.

% \hang{sec:logging phase}
% \hang{sec:replay}
% \hang{sec:root-causing}
% \hang{each section: what are the capabilities, what are the extendability, use a image classifcation thorughout the paper, use visual images, talk about how developer can extends}
% \vspace{-2mm}
% \subsection{EdgeML Monitor}
% \label{sec:monitor}

% \hang{move this figure to mldiff, replace with an running exmaple thoughtout the paper }

% \hang{input output view, what can be used to do what at each stage}

% \hang{start with an example, separate figure, running through example, model data source, }

\vspace{-1mm}
\subsection{API for Instrumentation and Logging}
\label{sec:api}
\vspace{-1mm}

To enable instrumentation on both the edge pipeline as well as the reference pipeline (\secref{sec:ref_pipeline}), \sysname needs to handle the heterogeneity of edge devices. Therefore, \sysname provides three sets of APIs: a cross-platform edge API (\eg c++ for Android, iOS, EdgeTPU), a platform specific API (\eg Java for Android), and an API for reference pipelines (Python). All the APIs follow the same data model (detailed below), focusing on three types of telemetry data.

As a concrete example, to invoke \sysname in a Tensorflow Lite app, a developer may write (in C++):
\begin{lstlisting}[language=C++]
MLEXray->on_inf_start();
TfLiteStatus s = m_interpreter->Invoke();
MLEXray->on_inf_stop(&m_interpreter);
\end{lstlisting}
\vspace{-2mm}
% To log the peripheral sensors during the camera shuttering, the developer may write (in Java)\footnote{Synchronous Android Camera2 API will not trigger camera shutter until ``onImageAvailable'' function exits.}:
% \begin{lstlisting}[language=Java]
% void onImageAvailable(ImageReader reader) {
%     MLEXray.on_sensor_stop();
%     // process image ...
%     MLEXray.on_sensor_start();}
% \end{lstlisting}
% \vspace{-2mm}
Users can also use \sysname APIs to log the input and output of any custom functions and peripharal sensors (see appendix \secref{sec:appd_design}). For example, to verify if the channels are extracted correctly, users can log the output of the extraction function from the edge pipeline as well as the reference pipeline (denoted as {\small\texttt{edge\_out}}, {\small\texttt{ref\_out}}).

% \subsection{Data Model and Assertion Functions}
% \label{sec:data_model}

% \vspace{-2mm}

We now formulate the data model for logging and give an example of designing assertion functions using the data. 

\parab{Data model.} The API focuses on logging three sets of data: 

% \parae{Input/Output:} including model input/output, per-layer input/output, the input/output of preprocessing and postprocessing functions, as well as the input/output of any user-defined functions in the inference pipeline.

% \parae{Performance metrics:} including ML inference end-to-end latency, per-layer latency, and memory footprint.

% \parae{Peripheral sensors:} including the sensors available on the devices, such as orientation, motion, ambient lighting conditions, \etc, that may affect input data quality and therefore degrade ML performance.
\vspace{-4mm}
\begin{itemize}[leftmargin=4mm]
    \item Input/Output: including model input/output, per-layer input/output, the input/output of preprocessing and postprocessing functions, as well as the input/output of any user-defined functions in the inference pipeline.
    
    \vspace{-2mm}
    
    \item Performance metrics: including ML inference end-to-end latency, per-layer latency, and memory footprint.
    
    \vspace{-2mm}
    
    \item Peripheral sensors: including the sensors available on the devices, such as orientation, motion, ambient lighting conditions, \etc, that may affect input data quality and therefore degrade ML performance.
\end{itemize}
\vspace{-4mm}
Each of these can be expressed as a key-value pair.

\parab{Assertion function.} An assertion function is an arbitrary function that can indicate whether a bug exists or an error happens in the deployed edge pipeline. For various purposes of validation, an assertion function can query different keys in the log of the same pipeline or same keys of two or more different pipelines. For example, the MobileNet model assumes an image input of RGB channel arrangement whereas Inception model assumes BGR. To validate the input channel against a correct reference pipeline for a particular model, an app developer may write (in Python):
% \begin{minipage}{\textwidth}
\begin{lstlisting}[language=Python]
def channel_assertion(edge_out, ref_out) {
    if not np.allclose(edge_out, ref_out):
        edge_out = cv2.cvtColor(edge_out, cv2.COLOR_BGR2RGB)
        if np.allclose(edge_out, ref_out):
            raise AssertionError('BGR->RGB')}
\end{lstlisting}
% \end{minipage}
% \hang{give picture example of bgr vs rgb image, or maybe later in eval}
\vspace{-2mm}
% \subsection{EdgeML Deployment Validation}

% We now describe the validation framework. \figref{fig:debuggingflowchart} shows a generic debugging flowchart. \sysname takes as input the model, the benchmark dataset, and the edge logs that the model produces on the dataset. It runs the reference pipeline using the model on the same dataset in a non-edge environment (\eg cloud, local workstation). Before doing deeper analysis, degraded accuracy is a first indicator of potential issues in the deployment process.  Then \sysname compares the logs from the mobile and cloud to identify and locate potential bugs and errors. \sysname consists of built-in assertions corresponding to discrepancies at different stages of the pipeline to analyze the potential reason. Finally, \sysname runs user-defined assertions to incorporate expert domain knowledge for root cause analysis.
\vspace{-1mm}
\subsection{Reference ML pipelines and Data Playback}
\label{sec:ref_pipeline}
\vspace{-1mm}
To identify issues of edge ML deployment, we need a correct reference pipeline as the baseline for comparison. However, establishing such baselines can be tricky.

From training an ML model to deployment, the actual model can have multiple versions throughout the entire process. Consider deploying a Tensorflow model to an android app as an example. First, the model presents as several checkpoint files during the training stage. Then, it is converted to one FlatBuffer file to run on the edge devices, while the conversion operation optimizes for inferences. Finally, out of the several quantization schemes\footnote{Such as \textit{quantization-aware training}, \textit{post-training dynamic-range quantization}, \textit{post-training float16 quantization}.}~\cite{jacob2017quantization}, we consider \textit{post-training full-integer quantization}, which converts model weights and biases from 32-bit floats to 8-bit and 32-bit integers, suitable for edge deployment. 

%  \hang{Jian, Pete can elaborate more on the para above}

To establish reference baselines, a first challenge is to have reference pipelines that can faithfully replay data using all these diverse versions of the model described above. For example, a comparison between a checkpoint model and a quantized model can potentially identify issues that happen during the conversion, while a comparison between a reference op and optimized op for the same quantized model can check if the optimized operation kernels are executed on the edge devices correctly. However, at the deployment and debugging stage, app developers 
% may not be model developers, and therefore may or 
may not have access to all versions of the model. 
% \sysname incorporates, as many as possible, widely used reference models for well-defined tasks such as classification, detection, and can accept user-defined pipelines as reference pipelines for extensibility.
This limits the baselines we can establish and may in turn restrict the debugging scope. 
% For example, if only the quantized model is available, \sysname can only 

A second challenge is that reference pipelines are often not available to app developers so they have to design them from scratch. 
% In this case, bugs due to the mismatch of the implicit assumptions between model developers and app developers are likely to happen again on the reference pipeline. 
Here is where the disconnect between model and app developers can cause mismatching assumptions again.
For example, the assumption of input preprocessing  (\eg resizing, normalization, \etc \secref{sec:background}) is unknown to app developers, which can degrade the reference pipeline performance, hence making the issue harder to uncover. 

To solve this, we designed a suite of correct reference pipelines for well-defined tasks and widely used models (\eg Mobilenet for image classification). These pipelines include preprocessing functions respecting model assumptions, reference model checkpoints, and various versions of the same model for edge devices. In addition to providing as many correct baselines as possible, \sysname also allows user-defined baselines. We will open-source the code to the community and build up the reference pipelines.

% \hang{emphasize this, open-source avaialbe to community}

% \input{fig_latex/debugflowchart}
\vspace{-1mm}
\subsection{Deployment Validation and Assertions}
\label{sec:validation}
\vspace{-1mm}
Given the logs from both the edge and the cloud reference pipeline, \sysname performs an initial accuracy validation (\figref{fig:sysarch_example}). 
% If there is a significant degradation, \sysname invokes MLDiff, which checks and visualizes per-layer output error (\eg mean squared error (MSE), mean absolute error (MAE)).
If there is a clear indicator of optimization issues (\eg accuracy drop), \sysname next checks per layer output differences to further diagnose the root cause. We evaluate, for each layer, the root-mean-square-error (rMSE) normalized by the layer output scale, denoted as $\widehat{rMSE} = {rMSE}/({max_{i}{(e_i)}-min_{i}{(e_i)}})$,
% ) 
% as follows:
% \begin{equation}
%     \widehat{rMSE} = \sqrt \frac{\sum_{i =1}^{N}{(e_i - r_i)^2}}{N} \times \frac{1}{max_{i}{(\mid e_i \mid)}-min_{i}{(\mid e_i \mid)}} 
% \end{equation}
where $\Vec{e}$ is the layer output vector. 
% of $N$ elements from the edge pipeline, and $\Vec{r}$ is that of the reference pipeline.
% The reason rMSE is chosen as an error measurement is 
Based on our past observation, rMSE normalized by scale tends to have a positive correlation with numerical deviation. Beyond rMSE, the \sysname framework allows easy extension to other error functions.
% , such as \textit{mean absolute error}.

% \hang{example pic for per-layer error}
% \hang{example pic for one layer vector compare}

$\widehat{rMSE}$ can be a fast indicator to locate potential issues. For example, given the same input to two different pipelines, a jump of $\widehat{rMSE}$ after a particular op can indicate an error in that op. If the error happens at the model input, the problem resides in the preprocessing functions. 
Common preprocessing errors (\secref{sec:background}) includes bugs in channel extraction, normalization scale, resizing functions, orientation, \etc 
% \figref{fig:common_preprocess_error} shows examples of these errors. 
These errors can easily degrade model accuracy by up to 30\% (\secref{sec:preprocess_bugs} details the impact of these bugs on classification accuracy). \sysname includes built-in assertions for each of these bugs, so that a simple automated validation can easily catch these bugs in user application code.
%
% \hang{insert pics here for various preprocessing bugs}
% \hang{generalize to audio classification task}
% \hang{move the definition of rmse here}
% \figref{fig:quant_issue} shows an example analysis of per-layer error for an image classification model. 
If the error happens at a specific layer in the model, this indicates ML operations that happen at that layer are error-prone. This can happen because of model quantization, or ML op optimization on particular edge devices. We identify two of these issues and discuss them in our evaluation (\secref{sec:quant_impact}). 

Similarly, following the pattern of validating per-layer output, \sysname can also perform per-layer latency validation. 
% (extracted using TFLite Benchmark~\cite{TFBenchmark}). 
% If the end-to-end latency is abnormal, 
\sysname can go over the latency of each layer and identify straggler layers in the model (see \secref{sec:latency_impact}). 

% \subsection{Extending Validation via User-defined Assertions}

\section{Evaluation}
\label{sec:eval}

\begin{figure*}
 \centering
  \includegraphics[width=1.8\columnwidth]{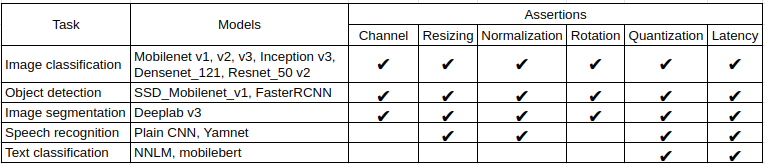}
  \vspace{-2mm}
  \caption{Summary of tasks, models and assertions in our evaluation. \sysname applies to a wide range of tasks and can identify deployment issues across multiple dimensions including input processing, quantization, and system performance such as abnormal latency.}
  \label{fig:res_summary}
  \vspace{-2mm}
\end{figure*}

% \subsection{Experimental Setup}
% \vspace{-2mm}
We evaluate \sysname by deploying various ML-based applications on mobile devices, referring to public deployment tutorials and open-source application repositories. To examine the efficacy of \sysname, we instrument these applications using \sysname APIs, and catch deployment issues by evaluating built-in as well as custom assertions. Our evaluation is designed to support the following claims:

% \begin{itemize}

% \item 
\parab{* \sysname catches a wide range of edge deployment issues  in industrial production systems across different ML applications.} Using ML applications of different sensor modalities (\eg image, audio, text), we show that \sysname can quickly identify a variety of input preprocessing bugs in the inference pipeline (\secref{sec:preprocess_bugs}), quantization issues of the model (\secref{sec:quant_impact}), and latency issues due to sub-optimal operations on target hardware (\secref{sec:latency_impact}). 
\vspace{-1mm}

% \item 
\parab{* \sysname is lightweight and easy to use.} We measure the efficiency and system overhead that \sysname incurs, \eg extra latency, memory usage, \etc We also show that \sysname significantly reduced the line-of-code (LoC) needed to catch deployment issues, lowering the barrier for debugging and demonstrating its capability of providing visibility into the deployment process.

% \hang{refer to production system again}
\vspace{-1mm}
% \item 
\parab{* \sysname improves app performance.} We quantify the impact of the identified deployment issues for various ML applications. Our evaluations show that, just by eradicating these issues found by assertions of a few LoC, \sysname can improve model performance on the edge by a significant margin. The evaluation raises awareness of these problems, and \sysname offers a systematic solution.

% \end{itemize}

Specifically, we write and deploy widely-used ML applications in everyday life, such as image classification (automated grocery store), object detection (surveillance, security camera), segmentation (autonomous driving), speech recognition (home assistant), text classification (sentiment analysis), \etc For each task, we use one or more off-the-shelf  pretrained models, such as different versions of MobileNet~\cite{howard2017mobilenets}, Densenet~\cite{Gao2017densenet}, SSD~\cite{liu2016ssd}, Deeplab~\cite{chen2017deeplab}, 
% Yamnet~\cite{yamnet}, 
MobileBert~\cite{sun2020mobilebert}. To evaluate \sysname, we deploy these apps on a Pixel 4 and Pixel 3 with 64-bit octa-core CPU (2.84GHz, 2.5GHz), and mobile GPUs (Adreno 640, 630). To benchmark performance, these apps are instrumented, using our APIs, in a way that they can accept data from an SD card in addition to the original sensor streams. We use standard task-specific datasets (\eg ImageNet~\cite{imagenet}, COCO~\cite{lin2014microsoft}, speech commands~\cite{speechcommandsv2}
% , IMDB movie reviews~\cite{imdb}, \etc) 
for the evaluation.

Our reference pipelines are adapted from the original training pipelines source code of each of the models evaluated. We run the reference pipelines
% , and the validation framework, 
on a workstation with a 2.2 GHz Intel Core i7 6-Core CPU and a GeForce 3070 GPU.

% \hang{talk more how bugs are introduced here}

\vspace{-1mm}
\subsection{\sysname can Identify a Wide Range of Deployment Issues across Various Tasks}
\label{sec:result_summary}
\vspace{-1mm}

\figref{fig:res_summary} shows a summary of our evaluation. As shown in the figure, \sysname is generic and can be applied to a variety of different models of  different tasks for edge deployment. These tasks range from image-based, to audio- and text-based applications. The architecture of models varies from convolutional neural networks~\cite{howard2017mobilenets} to language embeddings~\cite{bengio2003neural_nnlm} and transformers~\cite{devlin2018bert, vision_transformer}. The instrumentation and logging using \sysname APIs are universal around different models and their edge inference pipelines. Some of the assertions such as quantization validation and system metrics check (such as latency and memory usage) are also task and model agnostic. When it comes to task-specific assertions, for example, image preprocessing assertions, they are applicable to a range of tasks, such as classification, detection, segmentation, pose estimation, \etc These preprocessing assertions include, but are not limited to, the following (see details in \secref{sec:background}): validating channel extraction, resizing functions, normalization scale, as well as input orientation.  A few of them, such as normalization and resizing function validation, are also applicable to other input modalities of multi-dimensional arrays, such as audio spectrograms\footnote{One preprocessing function for audio waveform is to transform it into a spectrogram using Fast Fourier Transform (FFT).}. Our evaluation shows that \sysname is capable of catching these issues among all these different applications.  In the following sections, we show that \sysname is easy to use (\secref{sec:overhead}). Also, we detail the issues we found on some of these deployed inference pipelines, and quantify their impact (\secref{sec:preprocess_bugs}, \secref{sec:quant_impact}, \secref{sec:latency_impact})

\vspace{-1mm}
\subsection{\sysname is Light and Easy to Use}
\label{sec:overhead}
\vspace{-1mm}

In this section, we quantify the overhead of using \sysname to validate edge deployment. There are three steps that incur overhead: 1) app instrumentation and writing assertion functions, 2) extra run-time latency, memory, and storage, and 3) offline validation procedures and assertions.

\parab{App instrumentation and assertion functions.} Without \sysname, app developers have to write a significant chunk of code to get the inference logs and compose assertion functions. \tabref{tab:LOC_reduction} shows how much easier it is to use \sysname for both instrumentations as well as assertions, compared to writing everything from scratch. Extracting and asserting output from custom preprocessing functions is easier (25 LoC), \sysname abstracts these to under 5 LoC. It is harder to check and validate per-layer details. For example, asserting per-layer output can verify the correctness of model optimization and quantization, per-layer latency can verify the efficiency of optimized operation on different hardware. \sysname abstracts per-layer logging, log parsing, and per-layer metric comparisons, of which the code can easily go over 100 LoC. With \sysname, users only need up to 15 LoC to scrutinize per-layer details.

\begin{table}[!hbt]
\small
    \centering
    \begin{tabular}{c | cc | >{\bfseries} c | cc | >{\bfseries} c}
        & \multicolumn{6}{c}{Line of Code}\\
        \cline{2-7}
         Debugging 	& \multicolumn{3}{c|}{W/ \sysname} & \multicolumn{3}{c}{W/O \sysname} \\
         \cline{2-7}
	Target & Inst & Asrt &	Total &	Inst	& Asrt & Total\\
	\hline
	\hline
Preprocessing 
	&1&	3	&4&	18	&7	&25\\
Quantization	&4	&9	&13&	82&	183&	265\\
Lat. \& Mem.&	4&	4&	8&	14&	8&	22\\
Per-layer Lat.&	2&	6&	8&	14&	90&	104

    \end{tabular}
    \vspace{-2mm}
    \caption{Line-of-code (LoC) of instrumentation and assertion to write for each debugging target. With \sysname APIs, app instrumentation can be cut down to less than 5 LoC. The \sysname assertion interface abstracts away log parsing, making assertions functions simpler and easier to write. }
    \label{tab:LOC_reduction}
    \vspace{-2mm}
\end{table}

% (eg. Channel, Resizing, Normalization, Rotation)
% \begin{figure*}
%  \centering
% \begin{minipage}{1\columnwidth}
%   \includegraphics[width=\columnwidth]{fig/image_preprocess_bug.png}
%   \caption{Image classification accuracy affected by preprocessing bugs: erroneous channel arrangement, different resizing functions, mismatching normalization scale, disoriented input.}
%   \label{fig:image_classficiation_preprocess_bugs}
% \end{minipage}
% \hfill
% \begin{minipage}{0.46\columnwidth}
%   \includegraphics[width=\columnwidth]{fig/detection_preprocess_bug.png}
%   \caption{Object detection mAP, affected by preprocessing bugs: erroneous channel arrangement, different resizing functions, mismatching normalization scale.}
%   \label{fig:detection_map_preprocess_bugs}
% \end{minipage}
% \hfill
% \begin{minipage}{0.46\columnwidth}
%   \includegraphics[width=\columnwidth]{fig/speech_preprocess_bug.png}
%   \caption{Speech recognition accuracy affected by spectrogram processing bugs.}
%   \label{fig:audio_classficiation_preprocess_bugs}
% \end{minipage}
  
% \end{figure*}

\begin{figure*}
 \centering
  \includegraphics[width=0.85\textwidth]{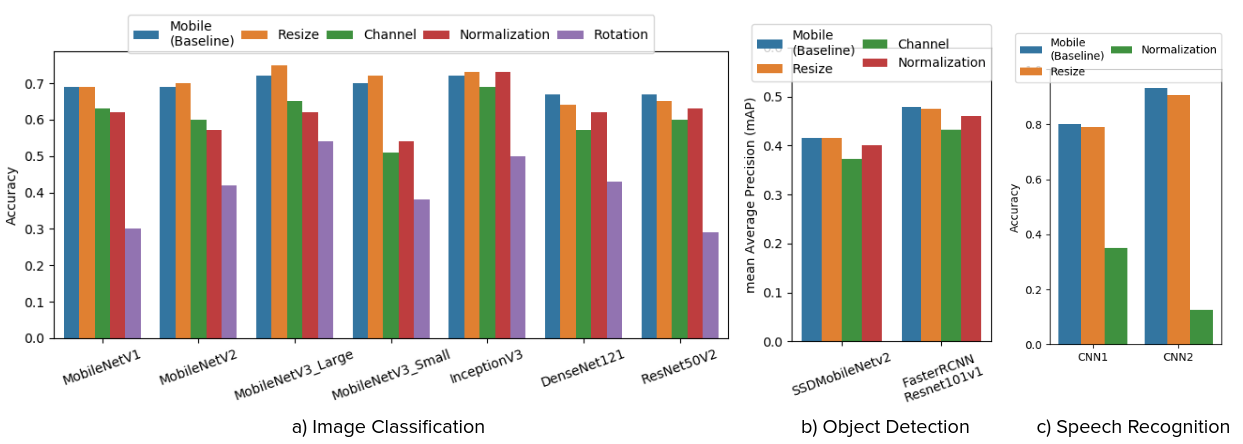}
  \vspace{-4mm}
  \caption{ML application performance across three different tasks degraded by preprocessing bugs found in industry deployment: erroneous channel arrangement, different resizing functions, mismatching normalization scale, and disoriented input.
%   , including models for a) Image classification, b) object detection, c) speech recognition.
  }
  \label{fig:image_classficiation_preprocess_bugs}
  \vspace{-4mm}
\end{figure*}

% \hang{using quotes from email, in motivation, reemphasize again here in exp setup}

\parab{App run-time overhead.} We next ask how much overhead \sysname incurs during app run-time. 
% Run-time log can be used to check accuracy, monitor end-to-end latency, and memory usage, using real time input data streams. These are the first indicators to initiate fine-grained per-layer output validations and assertions.
% , which only need to run one or a few samples offline.  
\tabref{tab:runtime_overhead} shows the latency, memory, and storage needed to run an instrumented app with \sysname. We instrumented and deployed an image classification app on Android Phones. The app runs MobileNet v2 over 100 images from the ImageNet test split. By default, we use 4 threads to run the ML model. 
As shown in \tabref{tab:runtime_overhead}, \sysname run-time logging only incurs up to 3 ms (15\%) latency when using Adreno mobile GPUs. Latency overhead is negligible (2.3\%) when using CPU. It also consumes about 3.7MB memory. 
% The logs are saved in an external storage (\ie an sdcard). 
Run-time logs are saved on an SDcard, and are as small as 0.41 KB per frame. 

\begin{table}[!hbt]
\small
    \centering
    \begin{tabular}{c | cc  cc }
	& Lat (ms)	& Lat (ms) &Mem &	Disk\\
	& CPU only	& GPU enabled &(MB) &	(KB/Frm)\\
	\hline
	\hline
Pixel 4 &	128.2±6.1&		16.7±0.3 	&6.42  &	-\\
P4(Inst) &	129.6±5.0 	&	19.1±0.6 	&10.12 & 0.41\\
\hline
Pixel 3 & 157.0±6.7	&		28.4±0.4	& 9.26&	-\\
P3(Inst) &	158.3±7.3	&	30.0±0.5	& 12.37 & 0.41\\
    \end{tabular}
    \vspace{-2mm}
    \caption{Run-time instrumentation overhead: latency, memory, and external storage. Numbers are from deploying an image classification app using Mobilenet v2 on Pixel 4 and Pixel 3 phones.}
    \label{tab:runtime_overhead}
    \vspace{-2mm}
\end{table}

\parab{Validation and assertion overhead.} As shown in the debugging flow chart \figref{fig:debuggingflowchart}, when \sysname finds an indication of deployment issues (\eg accuracy degradation) from the run-time logs, it triggers a fine-grained offline validation. This step involves logging per-layer output from both the edge and the reference pipeline. Depending on the complexity of the model deployed, the log size and the latency can vary a lot. To show the difference, we pick models with increasing layers (from 92 to 429) and the number of parameters of different magnitudes (from 3.5M to 25M).

% \hang{add calling out roughly linear relation}

\begin{table}[!hbt]
\small
    \centering
    \vspace{-2mm}
    \begin{tabular}{c | cc|ccc }
		& Layer & Param&	Lat &	Mem	& Disk\\
		& \#  &	\#&	(sec) &	(MB) 	& (MB)\\
		\hline
		\hline
Mobilenet v1&	92	&4.2M&	14 &	2&	44\\
Mobilenet v2&	156&	3.5M&	16	&1	&51\\
Resnet50 v2	&192&	25.6M&	67	&24&	176\\
Inception v3&	313	&23.9M&	57	&46	&150\\
Densenet 121&	429	&8M&	70	&74	&177\\
    \end{tabular}
    \vspace{-2mm}
    \caption{Offline validation overhead: latency, memory, and external storage (quantized 8-bit integer model). }
    \label{tab:assertion_overhead_quant}
    \vspace{-2mm}
\end{table}

\tabref{tab:assertion_overhead_quant} shows the latency, memory and storage space needed for these models, in the quantized 8-bit integer version (see \tabref{tab:assertion_overhead} in appendix for  32-bit float version). The evaluation shows that logging per-layer details can incur 14 to over 100 seconds (linear to model complexity) on edge devices (\eg Pixel 4). The memory footprint of these models varies from a few to hundreds of MB. The per-layer logs range from 44 MB to over 400 MB. The logs from the reference pipelines of the same model are of the same magnitude. Fortunately, comparing these two logs takes only a few seconds on commodity workstations, which is negligible given the logging latency (two orders of magnitude longer).

% \hang{add MLDiff latency memory overhead if space permits}

% 2) replay faithfulness, especially the one correponding to the mobile pipeline

% 3) performance of assertions, being able to scale

% easy to use, low overhead

% Raises awareness of the potential bugs due to implicit assumptions, bridges the gap between ML devs and App devs

% warm up, run single test image for per-layer io logging

% runtime logging, latency peripheral, model end-to-end IO

% \hang{consider swaping \secref{sec:bug_impact} and \secref{sec:overhead}}
% \vspace{-2mm}
% \subsection{Identified Issues and Impact}
% \label{sec:bug_impact}

% We describe the deployment issues \sysname identified and quantify their impact to application performance. 
\vspace{-1mm}
\subsection{Preprocessing Bugs and Impact}
\label{sec:preprocess_bugs}
\vspace{-1mm}

ML applications usually involve a few standard, yet error-prone, preprocessing steps for sensor input (\secref{sec:background}),  including, but not limited to, channel extraction, resizing, numerical conversion, and orientation. \sysname catches bugs in these functions by comparing the output of these functions to correct reference processing pipelines using user-defined assertions. To benchmark the impact of these bugs, we instrument Android ML apps to take input data from external storage instead of sensor streams, and run the app over publicly available task-specific benchmark datasets. 

\figref{fig:image_classficiation_preprocess_bugs} (a) shows the image classification accuracy of various models when one erroneous preprocessing function is applied. These preprocessing bugs, ranked by the severity of impact on Top-1 accuracy, include using 1) a different resizing function (\textbf{\textit{Resize}}, orange), 2) an erroneous channel arrangement (\textbf{\textit{Channel}}, green), 3) a mismatching normalization scale (\textbf{\textit{Normalization}}, red), and 4) a disoriented input (\textbf{\textit{Rotation}}, purple). For all models, the \textit{baseline} (optimized 32-bit float model, denoted as \textbf{\textit{Mobile}}, blue) always uses the same preprocessing functions from their training pipeline (\eg resizing using area averaging, arranging channel order as RGB, normalizing input to [-1.0,1.0], and using the original image orientation for Mobilenet v2). Each bar \textit{independently} introduces one and only one erroneous preprocessing function without inheriting from other bars. 

% \hang{emphasize again these bugs are discovered from industrial/google}

Compared to the baseline:
a) using a different resizing function in the inference pipeline (\eg bilinear resampling vs. area averaging), the Top-1  accuracy can vary by 1-3\%;
% \hang{report MSE of these layers too}
%
b) using a different channel arrangement order (\eg RGB vs. BGR), the model accuracy can degrade by 7-19\%, depending on how much the models care about the structural features versus the absolute pixel values; 
c) using a different normalization scale (\eg [0.0,1.0] vs. [-1.0,1.0]), the model accuracy can drop by up to 20\% easily;
d) rotating the image by 90 degrees (to emulate a disoriented image captured from mobile devices) can immediately break the app with the most severe 21-39\% accuracy drop, even though most of these models are trained with data augmentations~\cite{shorten2019survey_data_aug}, such as random rotation and flipping.

\begin{figure*}
 \centering
 \begin{minipage}{0.66\columnwidth}
 \centering
  \includegraphics[width=\columnwidth]{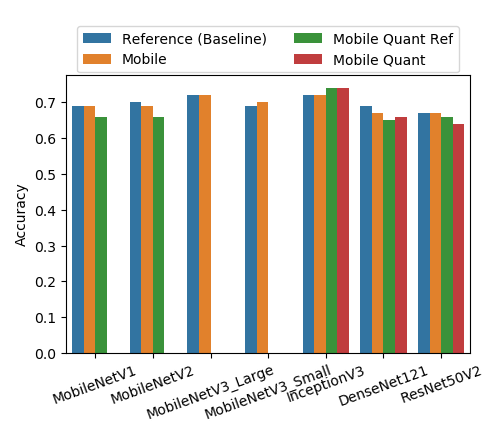}
  \vspace{-10mm}
  \caption{Model top-1 accuracy affected by optimization and quantization.}
  \label{fig:quant_model_accruacy}
%   \vspace{-3mm}
 \end{minipage}
 \hfill
 \begin{minipage}{1.37\columnwidth}
 \centering
 \includegraphics[width=1.05\columnwidth]{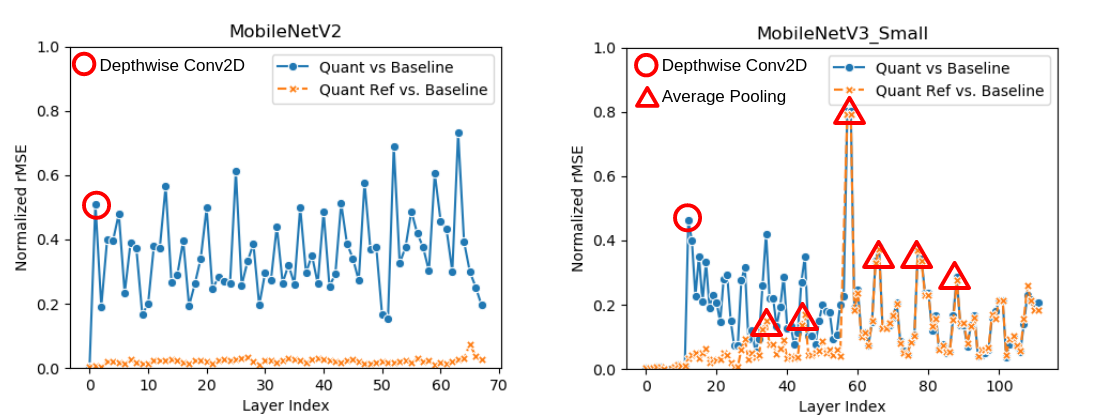}
 \vspace{-8mm}
  \caption{Normalized root-mean-square-error (rMSE) between quantized model vs. reference baseline for MobileNet v2 and v3.}
  \label{fig:quant_issue}
 \end{minipage}
 \vspace{-3mm}
\end{figure*}

% \hang{refer the github issues and comments from tf team}

% \hang{add other modality results}

\parae{Other tasks.}
We also evaluate the impact of preprocessing bugs on other ML applications, including object detection, segmentation, speech recognition, and text classification, \etc  Intuitively, the preprocessing bugs can affect other image-based applications. \figref{fig:image_classficiation_preprocess_bugs} (b) shows an example of degraded mean average precision of object detection models on COCO~\cite{lin2014microsoft} validation set for SSD~\cite{liu2016ssd} and FasterRCNN~\cite{ren2015faster}. Channel misarrangement and erroneous normalization can lower mAP by up to 4\%, while a different resizing function changes mAP by only 0.1\%. 
Preprocessing bugs can also impact tasks of different sensor modalities. \figref{fig:image_classficiation_preprocess_bugs} (c) shows lower audio recognition accuracy on a speech command dataset~\cite{speechcommandsv2}. In \figref{fig:image_classficiation_preprocess_bugs}, we use two speech models from different training pipelines~\cite{speech_tutorial}. Mismatching spectrogram normalization can significantly hurt these speech models.
We have also evaluated segmentation and text sentiment classification tasks, where the impact of these preprocessing bugs is less significant (see  Appendix \secref{sec:appd_result}).

\vspace{-1mm}
\subsection{Quantization Issues and Impact}
\label{sec:quant_impact}
\vspace{-1mm}

In order to validate model optimization and quantization, we leverage the per-layer details from \sysname logs. In addition to verifying the performance of the optimized model against the original one, \sysname also scrutinizes, layer by layer, to identify small output drifts or huge discrepancies. The former of which indicates arithmetic resolution issues, while the latter indicates incorrect operations. 

\figref{fig:quant_model_accruacy} shows the image classification accuracy of various models in increasingly optimized versions during the deployment process. For each model, here the baseline (\textbf{\textit{Reference}}, blue) is the original checkpoint format from the training pipeline. The other bars represent 1) an optimized 32-bit float model (\textbf{\textit{Mobile}}, orange), 2) a quantized 8-bit integer model using \textit{OpResolver}~\cite{OpResolver} (\textbf{\textit{Mobile Quant}}, red), a \textit{built-in operation resolver} that invokes an ``optimized kernel'' for production, and 3) the same fully quantized 8-bit integer model using \textit{RefOpResolver}~\cite{RefOpResolver} (\textbf{\textit{Mobile Quant Ref}}, green), a \textit{built-in reference operation resolver} that invokes a ``reference kernel'' for debugging (to rule out the possibility of issues caused by optimization\footnote{In TensorFlow Lite, a reference kernel is an easy-to-understand but inefficient implementation, usually in the form of naive loops in C/C++ without considering cache locality. An optimized kernel, on the other hand, can use assembly or intrinsics, fully utilizes cache locality, avoids branching, or adopts im2col \etc, hence making it hard to debug.}). Optimization for \textit{OpResolver} could cause small discrepancies on float models due to the non-associativity of floating point arithmetic. So any accuracy discrepancies in int8 fully-quantized model between builtin op and builtin reference op should be treated as a bug, which could mainly come from different \textit{overflow behavior} in the optimized kernel and the reference kernel. From a user perspective, the major difference between \textit{RefOpResolver} and \textit{OpResolver} is execution speed, and \sysname can leverage the two OpResolvers to provide debugging insights.
Beyond the two \textit{built-in op resolvers}, advanced users have the option to create their own OpResolver which could invoke their custom ops and kernels, which may or may not deviate from built-in behaviors.

We now summarize the results and dive deeper into the root causes.
Compared to the baseline: a) even without any quantization, running the converted 32-bit float model on mobile phones can already have a 1-2\% accuracy drop (\textbf{Baseline vs. Mobile}) (in that the specific arithmetic operations are not entirely the same), b) correctly executed quantized models using RefOpResolver can further incur $\pm 3\%$ accuracy change for MobileNet v1, v2, Inception v3, and Resnet 50 v2 (\textbf{Mobile vs. Mobile Quant Ref}). However, it can result in 0\% accuracy (MobileNetv3) with invalid or constant output. In cases where RefOpResolver does not have a significant accuracy drop (\eg MobileNet v1, v2), quantized models using optimized kernel can be error-prone to specific operations, yielding invalid or constant output (\textbf{Mobile Quant Ref vs. Mobile Quant}).

To reason about getting 0\% accuracy from quantized models on MobileNet (\figref{fig:quant_model_accruacy}), we show the per-layer diagnosis result of two examples (v2 and v3) that reveal different issues (b) and c) above. \figref{fig:quant_issue} depicts the normalized rMSE of each layer output when comparing Mobile Quant (blue), and Mobile Quant Ref (orange) against the baseline. The numbers are drawn from running a MobileNet v2 (left) and v3 (right) model on a Pixel 4. 

In the case of v2, using RefOpResolver, the $\widehat{rMSE}$ is always below 10\%, which explains the 1-3\% accuracy drop. However, the quantized model using the built-in op resolver shows a jump of rMSE at the second layer (red circle) of MobileNet v2 (also the 13th layer of v3). Further inspection of the model architecture reveals that these are DepthwiseConv2D layers, which indicates that the optimized op resolver has a bug on the depthwise convolution op.

Ruling out the factor of the op resolver, we take another look at the case of Mobile Quant Ref on MobileNet v3, which, despite the correct execution of depthwise convolution, still has 0\% accuracy. The right figure of  \figref{fig:quant_issue} shows that the normalized rMSE is having several peaks (red triangles) at layer 33, 44, 57, 64, 77, 87, \etc These peaks respond to the average pooling layer, which is an addition from v2 to v3, in each residual block. 

At this point, \sysname can confidently report the error-prone layers and direct the user to further inspect the op implementation of these layers. In fact, \sysname  is the first to uncover these two issues. Thanks to \sysname, they are now under the radar of the Tensorflow developers.

\vspace{-1mm}
\subsection{Sub-optimal Kernels and Latency}
\label{sec:latency_impact}
\vspace{-1mm}
Depending on run-time optimization, ML operations executed on different platforms can have drastically different latency. For example, although reference op resolver may guarantee a correct execution, it can incur over 200$\times$ longer latency on mobile devices. \tabref{tab:perlayerlat} summarizes \sysname total per-layer latency by layer type
% \footnote{
% For per-layer latency, \sysname builds on top of Tensorflow performance measurement: 
% \url{https://www.tensorflow.org/lite/performance/measurement}} 
of MobileNetv2.

\begin{table}[!hbt]
\small
    \centering
    \begin{tabular}{l|cccc }
Layer &	Mobile&	Mobile&	Mobile	& Emulator(x86)	\\
 Type &	&	Quant&	Quant 	& Mobile \\
  (Count) &	(ms) &	(ms)&	 Ref (ms)	&(ms) \\
\hline
\hline

D-Conv(17)&	\textbf{95.4}&	22.7&	2885.2&	120.0	\\
Conv(35)	&	23.5&	\textbf{32.3}	&\textbf{18662.3}&	\textbf{1409.8}\\
FC(1)		&7.4&	7.1&	7.0&	71.2\\
Mean(1)	&	6.1&	5.6&	5.0&	2.5\\
Pad(4)		&1.6	&18.7&	60.8&	104.8\\
Add(10)	&	1.5&	7.7&	99.8&	7.0\\
Softmax(1)		&0.4	&0.0	&0.0&	0.2	\\
Quantize(1)& -	&		3.3&	0.7&-	\\
\hline
Total &	136.26&	97.816	&21721.2&	1715.7\\
    \end{tabular}
    \vspace{-2mm}
    \caption{Latency by layer type of MobileNetv2 executed on Pixel 4 and android emulator for Pixel 4.}
    \label{tab:perlayerlat}
    \vspace{-4mm}
\end{table}

% \begin{table}[!hbt]
% \small
%     \centering
%     \begin{tabular}{l|ccccc }
% Layer &	Mobile&	Mobile&	Mobile	&Emulator& Emulator	\\
%  Type &	&	Quant&	Quant 	& Mobile& Mobile \\
%   (Count) &	(ms) &	(ms)&	 Ref (ms)	&(ms) & Quant (ms)\\
% \hline
% \hline

% D-Conv(17)&	\textbf{95.4}&	22.7&	2885.2&	120.0	&162.9\\
% Conv(35)	&	23.5&	\textbf{32.3}	&\textbf{18662.3}&	\textbf{1409.8}&	\textbf{12014.6}\\
% FC(1)		&7.4&	7.1&	7.0&	71.2&	7.7\\
% Mean(1)	&	6.1&	5.6&	5.0&	2.5&	2.3\\
% Pad(4)		&1.6	&18.7&	60.8&	104.8&11.9\\
% Add(10)	&	1.5&	7.7&	99.8&	7.0&	11.2\\
% Softmax(1)		&0.4	&0.0	&0.0&	0.2	& 0.03\\
% Quantize(1)& -	&		3.3&	0.7&-	&3.7\\
% \hline
% Total &	136.26&	97.816	&21721.2&	1715.7&	12214.6\\
%     \end{tabular}
%     \caption{Latency by layer type of MobileNetv2 executed on Pixel 4 and android emulator for Pixel 4.}
%     \label{tab:perlayerlat}
% \end{table}

The latency differences of each layer type  are very interesting. a) Quantized 2D convolution layer is slower than the unquantized one using optimized op resolver. b) Each depth-wise 2D convolution is 8$\times$ heavier than normal 2D convolution layer on 32-bit float model, whereas quantized depthwise convolution is faster than normal convolution layer. c) Unoptimized reference op resolve is three orders of magnitude slower than the optimized one: it significantly slows down convolution, depth-wise convolution, padding and addition. d) operations are ARM-specific, which cannot benefit on x86 emulators. The 32-bit float model on an emulator has comparable depth-wise convolution layer latency, but is 44x slower on normal convolution layers. It remains a challenge to faithfully emulate edge device performance metrics on platforms with more compute but different architecture, because op optimizations are architecture-specific.

% \subsubsection{Input Data Quality}
% \hang{AV camera angle, if time permits}
% \input{section/5_discussion}
\section{Related Work}
\label{sec:related}

\parab{ML inference at the edge.} Recent years have seen growing demand for running ML applications with latency, bandwidth, and privacy constraints on edge devices. These edge devices are equipped with special low-power hardware (\eg Qualcomm Adreno GPUs~\cite{adreno}, Coral edge TPUs~\cite{edge_tpu}, Intel VPUs~\cite{vpu}, micro-controllers~\cite{arduino} and DSPs). To push ML to the edge, the research and industry community have put significant effort (\eg TinyML~\cite{warden2019tinyml}, Tflite~\cite{lee2019ondevice}, TfliteMicro~\cite{MLSYS2021_d2ddea18}, PyTorchMobile~\cite{torch_mobile}) in optimizing the ML execution on these heterogeneous hardware devices. 

% To push ML to the  Tiny ML~\cite{warden2019tinyml}, Tflite on mobile GPUs~\cite{lee2019ondevice}, Tflite micro~\cite{MLSYS2021_d2ddea18}, 

\parab{ML profiling and benchmarks.} Previous work has focused on profiling ML training~\cite{mattson2019mlperf_train}, inference performance~\cite{reddi2020mlperf_inf} in the cloud~\cite{paradnn}. Recent work looks at profiling kernel operations (DeepBench~\cite{deepbench}) of different ML frameworks~\cite{luo2020comparison} with respect to different  hardware~\cite{MLSYS2021_02522a2b,  zhang2021nn-meter, tao2017benchip} at the edge. However, there is very little work on validating and debugging the deployment process.

% MLPerf, training~\cite{mattson2019mlperf_train}, inference~\cite{reddi2020mlperf_inf}, mobile inference~\cite{reddi2020mlperf_mobile_inf} 

% benchmarking ML framework~\cite{luo2020comparison}

% nnMeter mobisys paper~\cite{zhang2021nn-meter}
% Yunxin MLsys paper~\cite{MLSYS2021_02522a2b}

% NN characterization or benchmarking works before, such as BenchIP (Tao et al., 2018), DeepBench (Baidu, 2020), MLPerf (Reddi et al., 2020), ParaDnn (Wang et al., 2020a) and others (Gao et al., 2020; Bianco et al., 2018; Zhang et al., 2019; Turner et al., 2018; Hadidi et al., 2019; Wu et al., 2019b),

\parab{ML Validation.} Recent work raises the awareness of model performance validation by introducing model assertions~\cite{kang2018model_assertion} to check inconsistent outputs from models, or being run across different devices~\cite{MLSYS2021_b53b3a3d}. Our work looks beyond the output and provides visibility inside the model, and focuses on the system issues of edge deployment~\cite{paleyes2021challenges}.

% \hang{Pete will take a cut.}

% snorkel paper
% smol daniel kang
\section{Conclusion}
In this work, we introduce \sysname, a validation framework for edge ML deployment. \sysname enables app developers to catch complicated deployment issues just by writing a few lines of instrumentation and assertion code. We implemented \sysname as a suite of multi-lingual instrumentation APIs and an end-to-end deployment validation library, which will be open-sourced. Using different ML models for image, audio, and text-based applications, we showed that \sysname can help catch a variety of issues including preprocessing bugs, model optimization and quantization issues, and suboptimal kernel execution. Eradicating these issues can substantially improve edge application accuracy and latency. Code and APIs will be open-sourced to the community as a multi-lingual instrumentation library and a Python deployment validation library.
\label{sec:concl}

\balance
\newpage
\bibliography{references}

\begin{thebibliography}{51}
\providecommand{\natexlab}[1]{#1}
\providecommand{\url}[1]{\texttt{#1}}
\expandafter\ifx\csname urlstyle\endcsname\relax
  \providecommand{\doi}[1]{doi: #1}\else
  \providecommand{\doi}{doi: \begingroup \urlstyle{rm}\Url}\fi

\bibitem[OpR()]{OpResolver}
Tensorflow lite built-in op resolver.
\newblock URL
  \url{https://github.com/tensorflow/tensorflow/blob/master/tensorflow/lite/kernels/register.h}.

\bibitem[Ref()]{RefOpResolver}
Tensorflow lite built-in reference op resolver.
\newblock URL
  \url{https://github.com/tensorflow/tensorflow/blob/master/tensorflow/lite/kernels/register_ref.h}.

\bibitem[adr()]{adreno}
Qualcomm adreno mobile gpu.
\newblock URL \url{https://www.qualcomm.com/products/features/adreno}.

\bibitem[ard()]{arduino}
Arduino platform.
\newblock URL \url{https://www.arduino.cc/}.

\bibitem[dee()]{deepbench}
Benchmarking deep learning operations on different hardware.
\newblock URL \url{https://github.com/ baidu-research/DeepBench}.

\bibitem[edg()]{edge_tpu}
Intel vpu.
\newblock URL \url{https://cloud.google.com/edge-tpu}.

\bibitem[ker()]{keras_app}
Tensorflow keras application module: tf.keras.applications.
\newblock URL
  \url{https://www.tensorflow.org/api_docs/python/tf/keras/applications}.

\bibitem[spe()]{speech_tutorial}
Simple audio recognition tutorial.
\newblock URL \url{https://www.tensorflow.org/tutorials/audio/simple_audio}.

\bibitem[tfm()]{tfmodelquant}
Tensorflow model optimization.
\newblock URL \url{https://www.tensorflow.org/model_optimization/guide}.

\bibitem[tor()]{torch_mobile}
Pytorch mobile.
\newblock URL \url{https://pytorch.org/mobile/}.

\bibitem[vpu()]{vpu}
Intel vpu.
\newblock URL
  \url{https://www.intel.com/content/www/us/en/products/details/processors/movidius-vpu.html}.

\bibitem[yam()]{yamnet}
Yamnet: a speech recognition model.
\newblock URL \url{https://tfhub.dev/google/yamnet/1}.

\bibitem[Bengio et~al.(2003)Bengio, Ducharme, Vincent, and
  Janvin]{bengio2003neural_nnlm}
Bengio, Y., Ducharme, R., Vincent, P., and Janvin, C.
\newblock A neural probabilistic language model.
\newblock \emph{The journal of machine learning research}, 3:\penalty0
  1137--1155, 2003.

\bibitem[Chen et~al.(2017)Chen, Papandreou, Kokkinos, Murphy, and
  Yuille]{chen2017deeplab}
Chen, L.-C., Papandreou, G., Kokkinos, I., Murphy, K., and Yuille, A.~L.
\newblock Deeplab: Semantic image segmentation with deep convolutional nets,
  atrous convolution, and fully connected crfs.
\newblock \emph{IEEE transactions on pattern analysis and machine
  intelligence}, 40\penalty0 (4):\penalty0 834--848, 2017.

\bibitem[Cidon et~al.(2021)Cidon, Pergament, Asgar, Cidon, and
  Katti]{MLSYS2021_b53b3a3d}
Cidon, E., Pergament, E., Asgar, Z., Cidon, A., and Katti, S.
\newblock Characterizing and taming model instability across edge devices.
\newblock In Smola, A., Dimakis, A., and Stoica, I. (eds.), \emph{Proceedings
  of Machine Learning and Systems}, volume~3, pp.\  624--636, 2021.
\newblock URL
  \url{https://proceedings.mlsys.org/paper/2021/file/b53b3a3d6ab90ce0268229151c9bde11-Paper.pdf}.

\bibitem[David et~al.(2021)David, Duke, Jain, Janapa~Reddi, Jeffries, Li,
  Kreeger, Nappier, Natraj, Wang, Warden, and Rhodes]{MLSYS2021_d2ddea18}
David, R., Duke, J., Jain, A., Janapa~Reddi, V., Jeffries, N., Li, J., Kreeger,
  N., Nappier, I., Natraj, M., Wang, T., Warden, P., and Rhodes, R.
\newblock Tensorflow lite micro: Embedded machine learning for tinyml systems,
  2021.

\bibitem[Deng et~al.(2009)Deng, Dong, Socher, Li, Li, and Fei-Fei]{imagenet}
Deng, J., Dong, W., Socher, R., Li, L.-J., Li, K., and Fei-Fei, L.
\newblock Imagenet: A large-scale hierarchical image database.
\newblock In \emph{2009 IEEE Conference on Computer Vision and Pattern
  Recognition}, pp.\  248--255, 2009.
\newblock \doi{10.1109/CVPR.2009.5206848}.

\bibitem[Devlin et~al.(2018)Devlin, Chang, Lee, and Toutanova]{devlin2018bert}
Devlin, J., Chang, M.-W., Lee, K., and Toutanova, K.
\newblock Bert: Pre-training of deep bidirectional transformers for language
  understanding.
\newblock \emph{arXiv preprint arXiv:1810.04805}, 2018.

\bibitem[Dosovitskiy et~al.(2020)Dosovitskiy, Beyer, Kolesnikov, Weissenborn,
  Zhai, Unterthiner, Dehghani, Minderer, Heigold, Gelly,
  et~al.]{vision_transformer}
Dosovitskiy, A., Beyer, L., Kolesnikov, A., Weissenborn, D., Zhai, X.,
  Unterthiner, T., Dehghani, M., Minderer, M., Heigold, G., Gelly, S., et~al.
\newblock An image is worth 16x16 words: Transformers for image recognition at
  scale.
\newblock \emph{arXiv preprint arXiv:2010.11929}, 2020.

\bibitem[Han et~al.(2016)Han, Mao, and Dally]{han2016deep}
Han, S., Mao, H., and Dally, W.~J.
\newblock Deep compression: Compressing deep neural networks with pruning,
  trained quantization and huffman coding, 2016.

\bibitem[Howard et~al.(2017)Howard, Zhu, Chen, Kalenichenko, Wang, Weyand,
  Andreetto, and Adam]{howard2017mobilenets}
Howard, A.~G., Zhu, M., Chen, B., Kalenichenko, D., Wang, W., Weyand, T.,
  Andreetto, M., and Adam, H.
\newblock Mobilenets: Efficient convolutional neural networks for mobile vision
  applications.
\newblock \emph{arXiv preprint arXiv:1704.04861}, 2017.

\bibitem[Huang et~al.(2017)Huang, Liu, Van Der~Maaten, and
  Weinberger]{Gao2017densenet}
Huang, G., Liu, Z., Van Der~Maaten, L., and Weinberger, K.~Q.
\newblock Densely connected convolutional networks.
\newblock In \emph{2017 IEEE Conference on Computer Vision and Pattern
  Recognition (CVPR)}, pp.\  2261--2269, 2017.
\newblock \doi{10.1109/CVPR.2017.243}.

\bibitem[Jacob et~al.(2017)Jacob, Kligys, Chen, Zhu, Tang, Howard, Adam, and
  Kalenichenko]{jacob2017quantization}
Jacob, B., Kligys, S., Chen, B., Zhu, M., Tang, M., Howard, A., Adam, H., and
  Kalenichenko, D.
\newblock Quantization and training of neural networks for efficient
  integer-arithmetic-only inference, 2017.

\bibitem[Kang et~al.(2018)Kang, Raghavan, Bailis, and
  Zaharia]{kang2018model_assertion}
Kang, D., Raghavan, D., Bailis, P., and Zaharia, M.
\newblock Model assertions for debugging machine learning.
\newblock In \emph{NeurIPS MLSys Workshop}, 2018.

\bibitem[Krishnamoorthi(2018)]{krishnamoorthi2018quantizing}
Krishnamoorthi, R.
\newblock Quantizing deep convolutional networks for efficient inference: A
  whitepaper.
\newblock \emph{arXiv preprint arXiv:1806.08342}, 2018.

\bibitem[Lee et~al.(2019)Lee, Chirkov, Ignasheva, Pisarchyk, Shieh, Riccardi,
  Sarokin, Kulik, and Grundmann]{lee2019ondevice}
Lee, J., Chirkov, N., Ignasheva, E., Pisarchyk, Y., Shieh, M., Riccardi, F.,
  Sarokin, R., Kulik, A., and Grundmann, M.
\newblock On-device neural net inference with mobile gpus, 2019.

\bibitem[Li \& Alvarez(2021)Li and Alvarez]{Li2021RNN}
Li, J. and Alvarez, R.
\newblock On the quantization of recurrent neural networks.
\newblock \emph{arXiv preprint arXiv:2101.05453}, 2021.

\bibitem[Lin et~al.(2014)Lin, Maire, Belongie, Hays, Perona, Ramanan,
  Doll{\'a}r, and Zitnick]{lin2014microsoft}
Lin, T.-Y., Maire, M., Belongie, S., Hays, J., Perona, P., Ramanan, D.,
  Doll{\'a}r, P., and Zitnick, C.~L.
\newblock Microsoft coco: Common objects in context.
\newblock In \emph{European conference on computer vision}, pp.\  740--755.
  Springer, 2014.

\bibitem[Liu et~al.(2016)Liu, Anguelov, Erhan, Szegedy, Reed, Fu, and
  Berg]{liu2016ssd}
Liu, W., Anguelov, D., Erhan, D., Szegedy, C., Reed, S., Fu, C.-Y., and Berg,
  A.~C.
\newblock Ssd: Single shot multibox detector.
\newblock In \emph{European conference on computer vision}, pp.\  21--37.
  Springer, 2016.

\bibitem[Luo et~al.(2020)Luo, He, Zhan, Wang, Gao, and Dai]{luo2020comparison}
Luo, C., He, X., Zhan, J., Wang, L., Gao, W., and Dai, J.
\newblock Comparison and benchmarking of ai models and frameworks on mobile
  devices, 2020.

\bibitem[Maas et~al.(2011)Maas, Daly, Pham, Huang, Ng, and Potts]{imdb}
Maas, A.~L., Daly, R.~E., Pham, P.~T., Huang, D., Ng, A.~Y., and Potts, C.
\newblock Learning word vectors for sentiment analysis.
\newblock In \emph{Proceedings of the 49th Annual Meeting of the Association
  for Computational Linguistics: Human Language Technologies}, pp.\  142--150,
  Portland, Oregon, USA, June 2011. Association for Computational Linguistics.
\newblock URL \url{http://www.aclweb.org/anthology/P11-1015}.

\bibitem[Mattson et~al.(2019)Mattson, Cheng, Coleman, Diamos, Micikevicius,
  Patterson, Tang, Wei, Bailis, Bittorf, et~al.]{mattson2019mlperf_train}
Mattson, P., Cheng, C., Coleman, C., Diamos, G., Micikevicius, P., Patterson,
  D., Tang, H., Wei, G.-Y., Bailis, P., Bittorf, V., et~al.
\newblock Mlperf training benchmark.
\newblock \emph{arXiv preprint arXiv:1910.01500}, 2019.

\bibitem[Nagel et~al.(2021)Nagel, Fournarakis, Amjad, Bondarenko, van Baalen,
  and Blankevoort]{nagel2021white}
Nagel, M., Fournarakis, M., Amjad, R.~A., Bondarenko, Y., van Baalen, M., and
  Blankevoort, T.
\newblock A white paper on neural network quantization, 2021.

\bibitem[Paleyes et~al.(2021)Paleyes, Urma, and
  Lawrence]{paleyes2021challenges}
Paleyes, A., Urma, R.-G., and Lawrence, N.~D.
\newblock Challenges in deploying machine learning: a survey of case studies,
  2021.

\bibitem[Reddi et~al.(2020{\natexlab{a}})Reddi, Cheng, Kanter, Mattson,
  Schmuelling, Wu, Anderson, Breughe, Charlebois, Chou,
  et~al.]{reddi2020mlperf_inf}
Reddi, V.~J., Cheng, C., Kanter, D., Mattson, P., Schmuelling, G., Wu, C.-J.,
  Anderson, B., Breughe, M., Charlebois, M., Chou, W., et~al.
\newblock Mlperf inference benchmark.
\newblock In \emph{2020 ACM/IEEE 47th Annual International Symposium on
  Computer Architecture (ISCA)}, pp.\  446--459. IEEE, 2020{\natexlab{a}}.

\bibitem[Reddi et~al.(2020{\natexlab{b}})Reddi, Kanter, Mattson, Duke, Nguyen,
  Chukka, Shiring, Tan, Charlebois, Chou, et~al.]{reddi2020mlperf_mobile_inf}
Reddi, V.~J., Kanter, D., Mattson, P., Duke, J., Nguyen, T., Chukka, R.,
  Shiring, K., Tan, K.-S., Charlebois, M., Chou, W., et~al.
\newblock Mlperf mobile inference benchmark.
\newblock \emph{arXiv preprint arXiv:2012.02328}, 2020{\natexlab{b}}.

\bibitem[Ren et~al.(2015)Ren, He, Girshick, and Sun]{ren2015faster}
Ren, S., He, K., Girshick, R., and Sun, J.
\newblock Faster r-cnn: Towards real-time object detection with region proposal
  networks.
\newblock \emph{Advances in neural information processing systems},
  28:\penalty0 91--99, 2015.

\bibitem[Savsunenko(2018)]{resizing_func}
Savsunenko, O.
\newblock How tensorflow’s tf.image.resize stole 60 days of my life, 2018.
\newblock URL
  \url{https://medium.com/hackernoon/how-tensorflows-tf-image-resize-stole-60-days-of-my-life-aba5eb093f35}.

\bibitem[Shangguan et~al.(2019)Shangguan, Li, Liang, Alvarez, and
  McGraw]{Shangguan2019Speech}
Shangguan, Y., Li, J., Liang, Q., Alvarez, R., and McGraw, I.
\newblock Optimizing speech recognition for the edge.
\newblock \emph{arXiv preprint arXiv:1909.12408}, 2019.

\bibitem[Shorten \& Khoshgoftaar(2019)Shorten and
  Khoshgoftaar]{shorten2019survey_data_aug}
Shorten, C. and Khoshgoftaar, T.~M.
\newblock A survey on image data augmentation for deep learning.
\newblock \emph{Journal of Big Data}, 6\penalty0 (1):\penalty0 1--48, 2019.

\bibitem[Simonyan \& Zisserman(2014)Simonyan and
  Zisserman]{simonyan2014very_vgg}
Simonyan, K. and Zisserman, A.
\newblock Very deep convolutional networks for large-scale image recognition.
\newblock \emph{arXiv preprint arXiv:1409.1556}, 2014.

\bibitem[Sun et~al.(2020)Sun, Yu, Song, Liu, Yang, and Zhou]{sun2020mobilebert}
Sun, Z., Yu, H., Song, X., Liu, R., Yang, Y., and Zhou, D.
\newblock Mobilebert: a compact task-agnostic bert for resource-limited
  devices.
\newblock \emph{arXiv preprint arXiv:2004.02984}, 2020.

\bibitem[Tan et~al.(2020)Tan, Pang, and Le]{tan2020efficientdet}
Tan, M., Pang, R., and Le, Q.~V.
\newblock Efficientdet: Scalable and efficient object detection, 2020.

\bibitem[Tang et~al.(2021)Tang, Han, Zhang, Cao, and Liu]{MLSYS2021_02522a2b}
Tang, X., Han, S., Zhang, L.~L., Cao, T., and Liu, Y.
\newblock To bridge neural network design and real-world performance: A
  behaviour study for neural networks.
\newblock In Smola, A., Dimakis, A., and Stoica, I. (eds.), \emph{Proceedings
  of Machine Learning and Systems}, volume~3, pp.\  21--37, 2021.

\bibitem[Tao et~al.(2017)Tao, Du, Guo, Lan, Zhang, Zhou, Xu, Liu, Liu, Tang,
  Rush, Chen, Liu, Chen, and Chen]{tao2017benchip}
Tao, J., Du, Z., Guo, Q., Lan, H., Zhang, L., Zhou, S., Xu, L., Liu, C., Liu,
  H., Tang, S., Rush, A., Chen, W., Liu, S., Chen, Y., and Chen, T.
\newblock Benchip: Benchmarking intelligence processors, 2017.

\bibitem[Wallace(1992)]{wallace1992jpeg}
Wallace, G.~K.
\newblock The jpeg still picture compression standard.
\newblock \emph{IEEE transactions on consumer electronics}, 38\penalty0
  (1):\penalty0 xviii--xxxiv, 1992.

\bibitem[Wang et~al.(2020)Wang, Wei, and Brooks]{paradnn}
Wang, Y.~E., Wei, G.-Y., and Brooks, D.
\newblock A systematic methodology for analysis of deep learning hardware and
  software platforms.
\newblock In \emph{The 3rd Conference on Machine Learning and Systems (MLSys)},
  2020.

\bibitem[{Warden}(2018)]{speechcommandsv2}
{Warden}, P.
\newblock {Speech Commands: A Dataset for Limited-Vocabulary Speech
  Recognition}.
\newblock \emph{ArXiv e-prints}, April 2018.
\newblock URL \url{https://arxiv.org/abs/1804.03209}.

\bibitem[Warden \& Situnayake(2019)Warden and Situnayake]{warden2019tinyml}
Warden, P. and Situnayake, D.
\newblock \emph{Tinyml: Machine learning with tensorflow lite on arduino and
  ultra-low-power microcontrollers}.
\newblock O'Reilly Media, 2019.

\bibitem[Zhang et~al.(2021)Zhang, Han, Wei, Zheng, Cao, Yang, and
  Liu]{zhang2021nn-meter}
Zhang, L., Han, S., Wei, J., Zheng, N., Cao, T., Yang, Y., and Liu, Y.
\newblock nn-meter: towards accurate latency prediction of deep-learning model
  inference on diverse edge devices.
\newblock In \emph{2021 International Conference on Mobile Systems,
  Applications, and Services}, pp.\  81--93. ACM, June 2021.

\bibitem[Zhou \& Tuzel(2018)Zhou and Tuzel]{zhou2018voxelnet}
Zhou, Y. and Tuzel, O.
\newblock Voxelnet: End-to-end learning for point cloud based 3d object
  detection.
\newblock In \emph{Proceedings of the IEEE conference on computer vision and
  pattern recognition}, pp.\  4490--4499, 2018.

\end{thebibliography}

\bibliographystyle{mlsys2021}

\newpage
\appendix

\section{Addtional Evaluation Results}
\label{sec:appd_result}

\parab{Deployment issues and impact on other tasks.} In addition to image classification, object detection, and speech recognition shown in \secref{sec:eval}, we have also deployed a text sentiment classification app using MobileBert~\cite{sun2020mobilebert}, and NNLM embeddings~\cite{bengio2003neural_nnlm} as well as image segmentation apps using Deeplab v3~\cite{chen2017deeplab}. We observe that even per-layer output values can be different in these models, the output accuracy is not significantly changed. For example, when NNLM takes raw texts versus the same text but with lower case, the embedding output is drastically different. However, the sentiment classification accuracy on the IMDB movie review dataset~\cite{imdb} is exactly the same. Further, some models, such as EfficientDet~\cite{tan2020efficientdet}, incorporate the input preprocessing functions into the model graph, which reduces the chance of having preprocessing bugs during deployment.

\parab{Offline validation overhead.}
We show the offline validation overhead (\tabref{tab:assertion_overhead}) for floating-point models. This is in correspondence to \tabref{tab:assertion_overhead_quant} in \secref{sec:overhead}
\begin{table}[!hbt]
\small
    \centering
    \begin{tabular}{c | cc|ccc }
		& Layer & Param&	Lat &	Mem	& Disk\\
		& \#  &	\#&	(sec) &	(MB) 	& (MB)\\
		\hline
		\hline
Mobilenet v1&	92	&4.2M&	21&	53&	87\\
Mobilenet v2&	156&	3.5M&	21&	6&	93\\
Resnet50 v2	&192&	25.6M&	104	&60&	478\\
Inception v3&	313	&23.9M&	85&	155&	333\\
Densenet 121&	429	&8M&	114	&298&	426\\
    \end{tabular}
    \caption{Offline validation overhead: latency, memory, and external storage (original 32 bit float model).}
    \label{tab:assertion_overhead}
\end{table}

\section{Additional Design Details}
\label{sec:appd_design}

\begin{figure}
 \centering
  \includegraphics[width=\columnwidth]{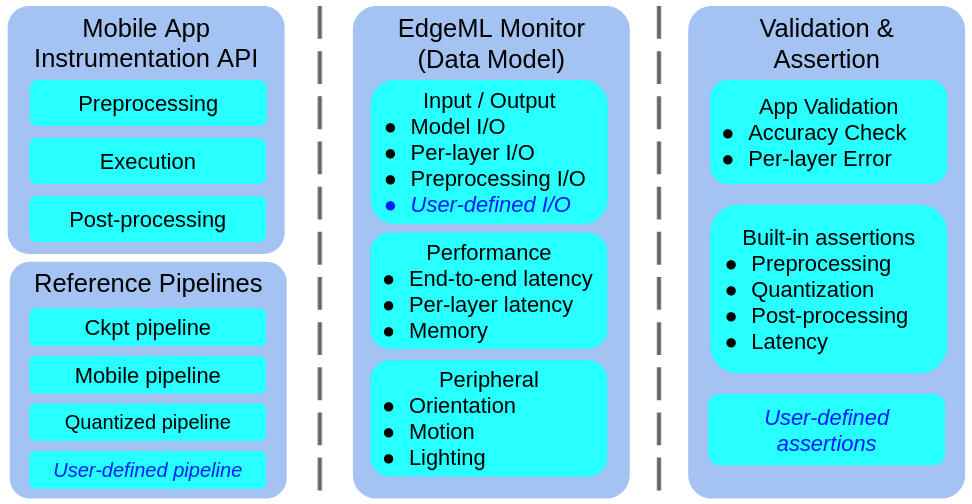}
  \caption{\sysname system diagram. \sysname includes app intrumentation APIs (\secref{sec:api}), reference ML inference pipelines (\secref{sec:ref_pipeline}), and deployment validation  framework (\secref{sec:validation}). Both the APIs and reference pipelines use an underlying EdgeML Monitor to log ML inference traces for validation and assertion. Users can customize logs and assertion functions, and provide user-defined pipelines as reference baselines.}
  \label{fig:sysarch}
  \vspace{-3mm}
\end{figure}

\figref{fig:sysarch} shows a system diagram. 
A general workflow is as follows. Both the instrumented mobile app (\secref{sec:api}) and the reference pipelines (\secref{sec:ref_pipeline}) will instantiate the EdgeML Monitor (\secref{sec:api}) to collect telemetry data of ML inferences. \sysname uses these data to compare and validate the mobile ML pipeline against the reference pipeline (\secref{sec:validation}). For well-defined tasks, such as image classification, object detection, audio recognition, \sysname provides built-in assertions on critical error-prone points (\eg preprocessing) where bugs are common. 

To invoke \sysname in Java, users can write:
\begin{lstlisting}[language=Java]
MLEXray.on_inf_start();
// invoke tflite...
MLEXray.on_inf_stop();
\end{lstlisting}
\vspace{-2mm}

To log the peripheral sensors during the camera shuttering, the developer may write (in Java)\footnote{Synchronous Android Camera2 API will not trigger camera shutter until ``onImageAvailable'' function exits.}:
\begin{lstlisting}[language=Java]
void onImageAvailable(ImageReader reader) {
    MLEXray.on_sensor_stop();
    // process image ...
    MLEXray.on_sensor_start();}
\end{lstlisting}

\end{document}